\RequirePackage{fix-cm}
\documentclass[twocolumn,epjc3]{svjour3}  
\smartqed  
\RequirePackage{graphicx}
\usepackage[dvipsnames]{xcolor}
\usepackage{gensymb}
\usepackage{multirow}
\newcommand{\tssc}[1]{\textsuperscript{#1}}
\journalname{Eur. Phys. J. C}
\begin{document}

\title{
Fast neutron production at the LNL Tandem
from the 
$^7$Li($^{14}$N,xn)X reaction
}


\author{
       Pablo Torres-Sánchez\thanksref{e1,addr1} \and
        Hans Th. J. Steiger\thanksref{e2,addr2, addr3} \and
        Pierfrancesco Mastinu\thanksref{addr4} \and
        Jeffery L. Wyss\thanksref{addr5,addr6} \and
        Lennard Kayser\thanksref{addr2} \and
        Luca Silvestrin\thanksref{addr5,addr7} \and
        Elizabeth Musacchio-González\thanksref{addr4} \and
        Matthias Raphael Stock\thanksref{addr2} \and
        David Dörflinger\thanksref{addr2} \and
        Ulrike Fahrendholz\thanksref{addr2} \and
        Gianfranco Prete\thanksref{addr7} \and
        Osvaldo Carletto\thanksref{addr4} \and
        Lothar Oberauer\thanksref{addr2} \and
        Ignacio Porras\thanksref{addr1}
}

\thankstext{e1}{e-mail: pablotorres@ugr.es}
\thankstext{e2}{e-mail: hsteiger@uni-mainz.de, hans.steiger@tum.de}

\institute{Dept. Atomic, Molecular and Nuclear Physics, University of Granada, Spain\label{addr1}
           \and
Technical University of Munich, TUM School of Natural Sciences, Physics Department, James-Franck-Str. 1, 85748 Garching, Germany \label{addr2}
           \and
           Cluster of Excellence PRISMA$^{+}$, Johannes Gutenberg University Mainz, Staudingerweg 9, 55128 Mainz, Germany \label{addr3}           
           \and
           Istituto Nazionale di Fisica Nucleare, Legnaro Division, viale dell'Università 2, 35020 Legnaro, Italy\label{addr4}
           \and
           Istituto Nazionale di Fisica Nucleare, Padova Division, via Marzolo 8, 35131 Padova, Italy,  \label{addr5}
           \and University of Cassino and Southern Lazio, Dipartimento di Ingegneria Civile e Meccanica, via G. Di Biasio 43, 03043 Cassino (FR), Italy\label{addr6}
           \and
           University of Padova, Dept. of Physics and Astronomy, via Marzolo 8, 35131 Padova, Italy\label{addr7} 
}

\date{Received: date / Accepted: date}

\maketitle

\begin{abstract}
Fast neutron beams are of relevance for many scientific and industrial applications. This paper explores fast neutron production using a TANDEM accelerator at the Legnaro National Laboratories, via an energetic ion beam (90 MeV $^{14}N$) onto a lithium target. The high energy models for nuclear collision of FLUKA foresee large neutron yields for reactions of this kind.
The experiment aimed at validating the expected neutron yields from FLUKA simulations, using two separate and independent set-ups: one based on the multi-foil activation technique, and the other on the time of flight technique, by using liquid scintillator detectors.

The results of the experiment show clear agreement of the measured spectra with the FLUKA simulations, both in the shape and the magnitude of the neutron flux at the measured positions. The neutron spectrum is centered around the 8 MeV range with mild tails, and a maximum neutron energy spanning up to 50 MeV.

These advantageous results provide a starting point in the development of fast neutron beams based on high energy ion beams from medium-sized accelerator facilities.

\end{abstract}

\section{Introduction}

    Neutron beams have important applications in many different fields.
    Most neutron beams are available at reactors and have cold or thermal spectra; such neutrons are uniquely suited to the study of the structure and dynamics of materials at the atomic level. 
On the other hand, few facilities can provide fast neutron beams ($E_n>1$ MeV); the access is limited and costs high. 

    Fast neutron sources are of importance not only for basic nuclear physics. The list of other fields of applications includes: dosimetry, neutron detector development, fast neutron oncology, radiation protection shielding materials for accelerator based oncology and Space missions, and the study of single-neutron induced effects in digital and power electronics.

    Tandem accelerators are still widely used in many small-medium size laboratories and could provide cost effective neutrons on existing beam lines. In addition to the direct applications to fundamental and applied physics mentioned above, important is also the possibility of training students in neutron physics.

    \section{Fast neutrons at the LNL Tandem}

    The Legnaro National Laboratories (LNL) of the Italian National Institute of Nuclear Physics is developing new sources of fast neutrons that would integrate the existing one at the CN Van de Graaff 6 MV accelerator and the one under design at the SPES 70 MeV proton cyclotron \cite{mastinu2020fast}\cite{Silvestrin2016}.
    To produce fast neutrons, we studied reactions involving heavy ion beams on light targets, as inverse kinematics can produce forwardly focused high energy neutrons in the laboratory reference frame.
    
Energetic ions beams are available at the Tandem-PIAVE-ALPI accelerator complex \cite{SIGNORINI198627}.  The experiment described below used the maximum energy available at the Tandem XTU accelerator; a followup experiment will use the PIAVE-ALPI heavy ion injector and superconducting linac to deliver higher energy beams. 

     As a solid target material, the lightest possible elements to be considered are lithium or beryllium. With regards to the beam and the performance of the Tandem, a balance between ion mass and available current gives nitrogen as the best compromise.
    The initial target choice is lithium, routinely used at the BELINA neutron facility at the CN accelerator; beryllium will soon be evaluated too.

\subsection{The experimental reaction}
The selected reaction $^7$Li($^{14}$N,xn)X
reaction  has a large $Q$-value (16741.137 keV); the kinematics, therefore, allows the production of fast neutrons with energies of tens of MeV, even at ion beam energies of less than 10 MeV/u.
The cross section and yield for this reaction are not available in literature. We used the Monte Carlo code FLUKA to study the neutron production spectra; the calculated FLUKA neutron spectra are based on theoretical models since no data are available for the reaction. 

Two independent measurement methods were used to measure the energy spectrum and validate the results of the simulations: the multi-activation foil technique and time-of-flight (TOF). 

The TOF part of this experiment is actually part of a broad research program for characterizing Liquid Scintillators developed for Neutrino Physics and Rare Event Astrophysics. 
In general, a detailed understanding of neutron signals in detectors is crucial to construct realistic simulations and predictions of natural phenomena; this is the immediate application of this new neutron source at LNL.

\subsection{Fluka Simulations}

Monte Carlo simulations of the nuclear reaction were performed using the models available in FLUKA \cite{Fluka}. Evaporation including heavy ion fragments and coalescence hadronic models were activated in the simulations. 

Figure 1 
shows the expected neutron yield at zero and 10 degrees, corresponding to the two positions where the fluence was measured. The position of the activation foil was tilted 10$\degree$ from the beam axis and was located 2.5 cm from the lithium target, covering a solid angle of 0.193 sr. The position of the TOF detector was located 4.56 m away from the lithium target in the forward (0$^\circ$) beam direction, covering a solid angle of $0.235\times 10^{-3}$ sr. The estimated yield at the activation position is lower than the one at the TOF detector position, confirming the strong forward angle dependence of the neutron production, typical of inverse kinematics reactions.
\begin{figure}
  \includegraphics[width=0.48\textwidth]{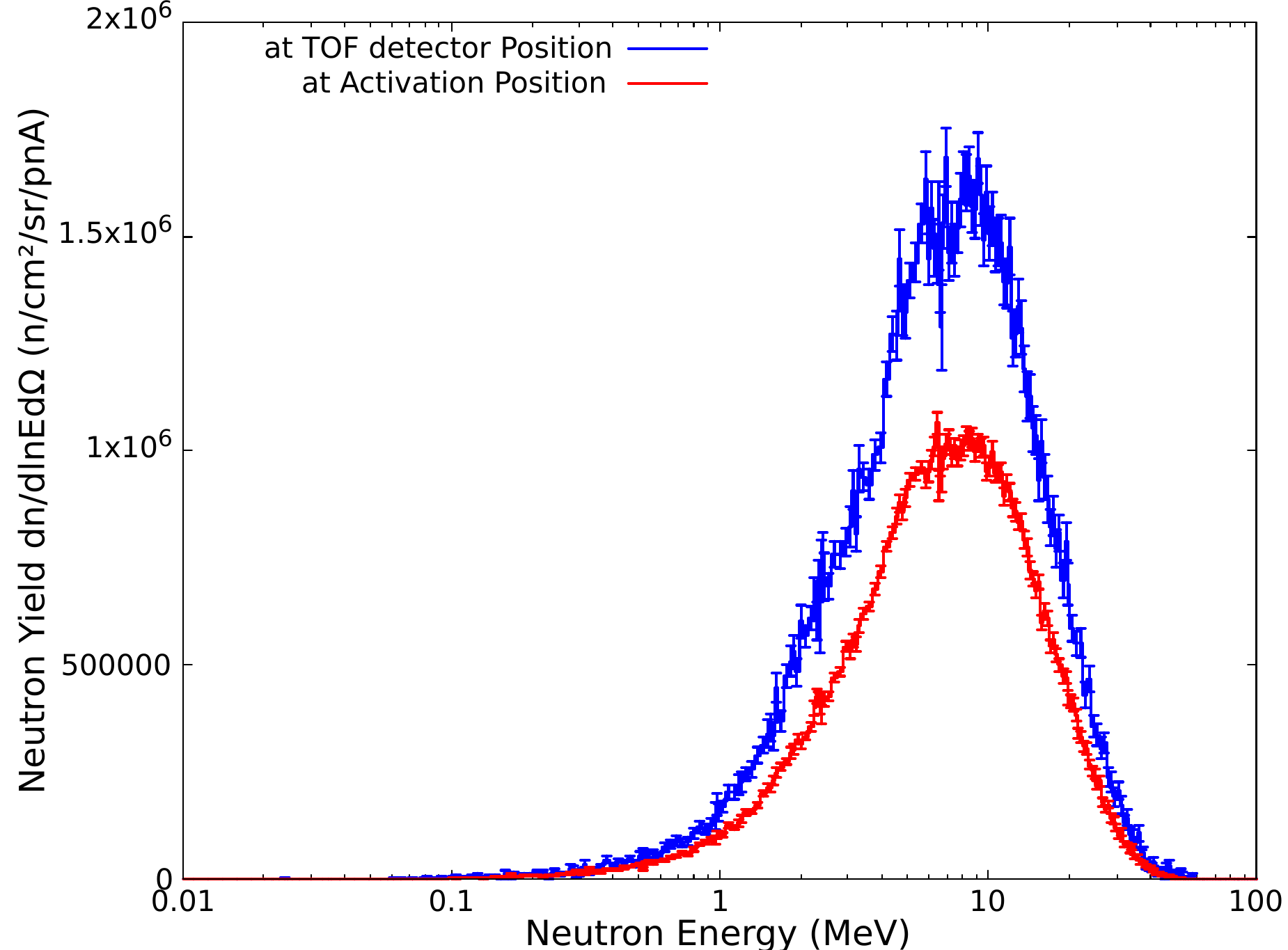}
\caption{FLUKA simulations of the neutron production at the two measurement positions. The integral yield towards the TOF detector in forward direction (blue distribution) is 1.6 times larger than that towards the activation foils at 10° (red).}
\label{fig:FlukaSimPlots}       
\end{figure}
%
\label{sec:materialsandmethods}

\subsection{The beam}
\begin{figure}[h]
  \includegraphics[width=0.48\textwidth]{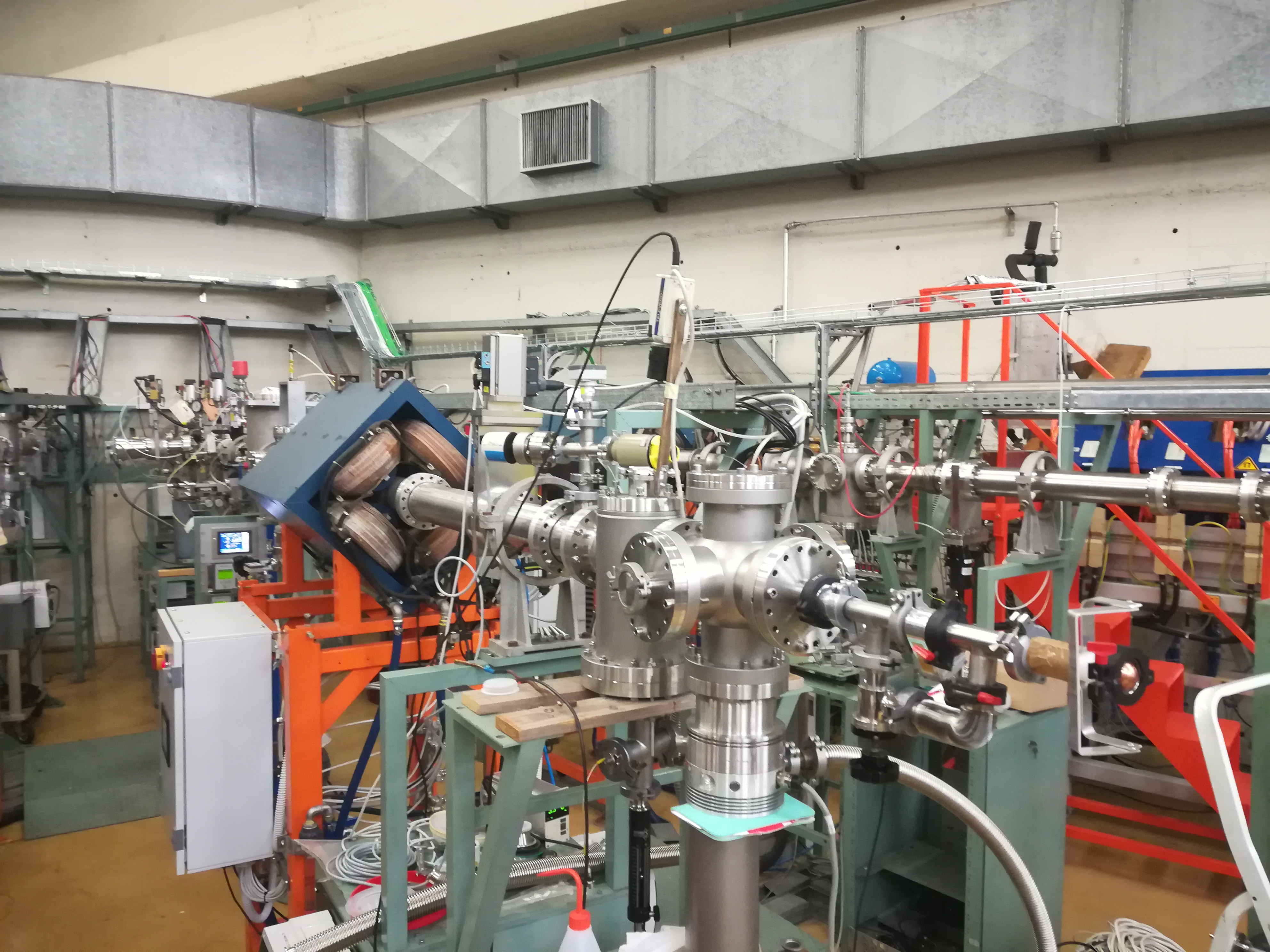}
\caption{The 30$^\circ$ beam line in Hall 1. The lithium target is encapsulated into a copper backing and beam stopper visible at the extreme right; a close-up is shown below in Figure 
\ref{fig:TargetActivationSamples}
}
\label{fig:beamline}       
\end{figure}

Negative BN molecules were injected into the Tandem, accelerated by a 13.7 MV terminal voltage, and dissociated by one carbon stripping foil to produce positive N ions that were then further accelerated. Magnetic Analysis selected $^{14}$N$^{+6}$ 90 MeV ions that were then transported along the 30$^\circ$ beam line in Hall 1 (Figure \ref{fig:beamline}) towards the lithium neutron production target. 

The beam was pulsed (800 ns repetition rate and 1 ns sigma pulse time width) in order to perform TOF measurements, thereby providing a cross-check of the simultaneous multi-foil activation measurement.

\subsection{The neutron production target}


\begin{figure}[h]
  \includegraphics[width=0.48\textwidth]{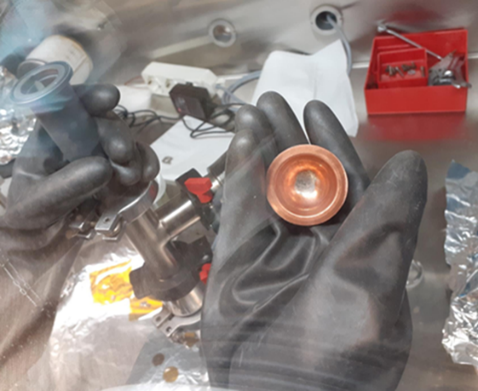}

\caption{Lithium pressed into a 300 $\mu$m thick copper backing and beam stopper. The production procedure, carried out inside an argon glovebox, is described in Section 2.4}
\label{fig:litium}       
\end{figure}

The target uses metal lithium encapsulated into a copper backing and beam stopper designed to have low mass and thus a low neutron spectra perturbability in forward direction.

 The target system is produced in an inert atmosphere inside an argon glovebox. The lithium is attached to the 300 µm thick Cu backing by pressing them between a male and a female mold, producing a lithium layer of the desired thickness, in this case, 200 $\mu$m (Figure \ref{fig:litium}). This technique is used regularly to 
 make thin lithium targets 
 used to produce neutrons at the CN accelerator. The copper backing has a hemispherical shape (see Figure \ref{fig:litium}) designed to minimize the material, and it is electrically insulated from the beam line. The current at the target is acquired directly from the copper backing. A two-stage tantalum collimator of 5 mm diameter is placed 10 cm upstream of the copper backing and the current delivered on it was also acquired.

\begin{figure}
  \includegraphics[width=0.48\textwidth]{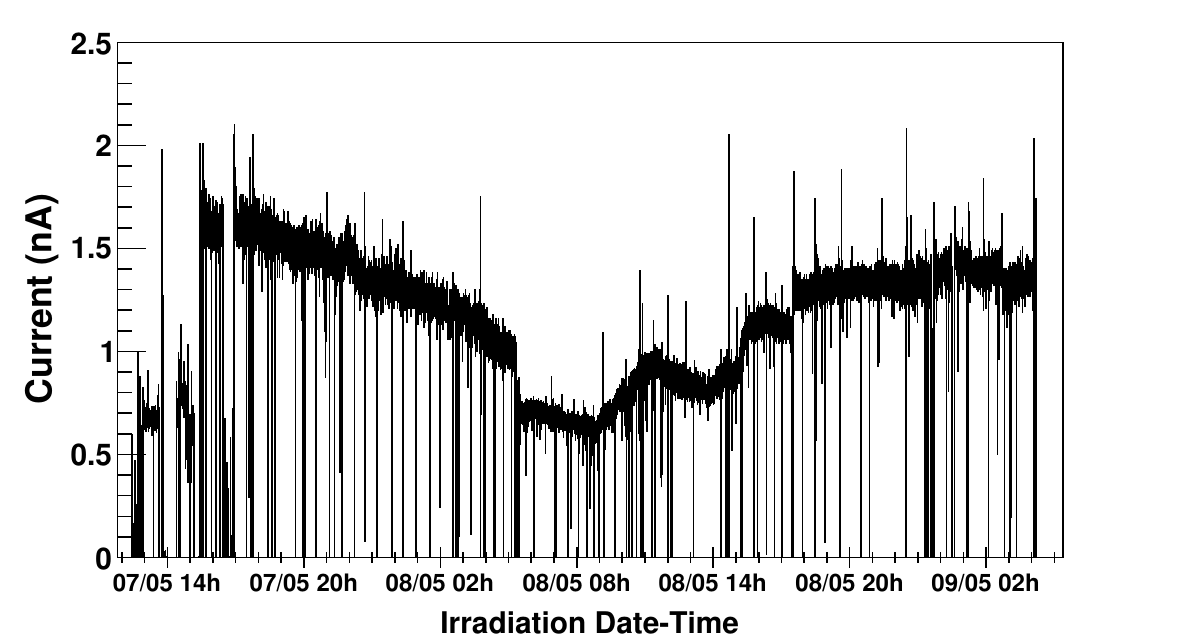}
\caption{Irradiation history of the measurement.}
\label{fig:IrradiationHistory}       
\end{figure}

The total charge accumulated during the experiment amounted to 0.153 mC. Figure \ref{fig:IrradiationHistory} shows the time-dependent ion current during the irradiation. 

\subsection{The multi-foil activation technique}

The activation technique is a well-established and very powerful method that can be used to measure neutron spectra. The main goal with this setup was to obtain broad spectral information including the low energy range below few MeV and specially to provide an absolute value for the yield at the measured position.

\subsubsection{Experimental setup}

For the multi-foil activation measurement, a series of samples especially sensitive to neutrons in the energy range of interest were selected(i.e., with large cross sections or with thresholds ranging from hundreds of keV to tens of MeV). A list of these activation foils and the reactions used are shown in Table \ref{tab:listsamples}. The activation foils were purchased from ShieldWerx \cite{shieldwerxfoils}. These samples were arranged in a stack configuration and were surrounded with two cadmium covers, to minimize the activation due to low energy neutrons. The particular ordering in the arrangement of samples was chosen based on the relative attenuation or scattering produced by each of the samples within the stack. These alterations in the neutron flux at each position were later corrected via dedicated Monte Carlo simulations using MCNP6.2 \cite{MCNP62}. Additional samples of Au were placed outside the Cd covers (Au-a sample) and in the last position (Au-c sample) and used for cross-checking the thermal contamination and possible scattering back from the experimental hall. However, these samples were not included in the unfolding of the neutron spectrum.


\begin{figure}[h]
  \includegraphics[width=0.48\textwidth]{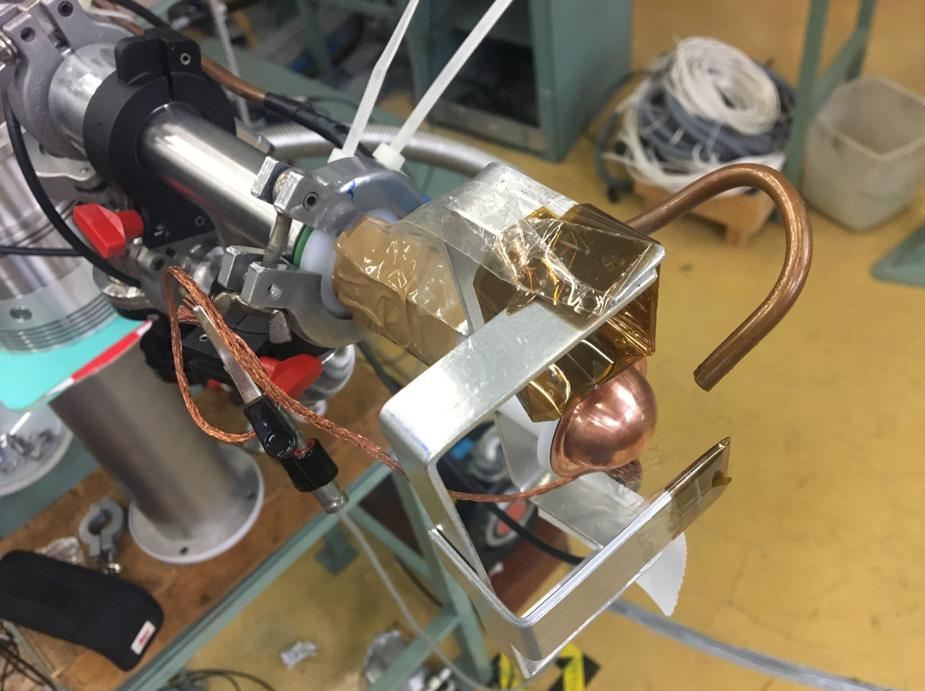}
  
  \vspace{1mm}
  
  \reflectbox{\includegraphics[width=0.48\textwidth]{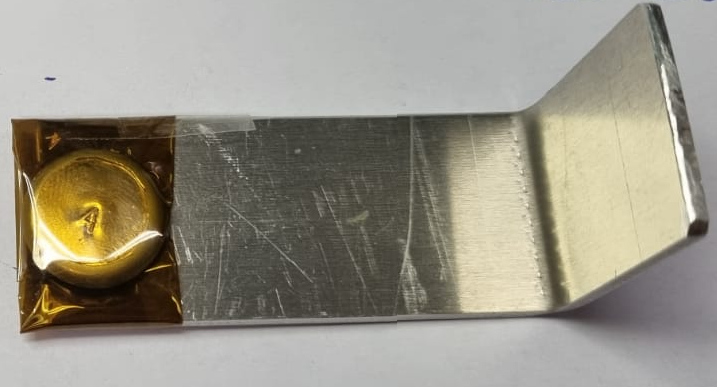}}
\caption{\textit{(Top)}: The final portion of the beam line with the lithium target installed (copper support), the air-cooling system for the target, and the aluminum support for the activation foils (aluminum mount).\\\textit{(Bottom)}: Detail of the aluminum holder used to keep the activation samples in front of the target.  The foils were secured using a kapton foil at the tip. }
\label{fig:TargetActivationSamples}       
\end{figure}

Figure \ref{fig:TargetActivationSamples} shows (top) the aluminum support and holder at its tilted position during the experiment; and (bottom) a detail of the sample holder. The stack of samples was mounted on an aluminum support and placed 2.5 cm from the center of the lithium target. The aluminum support was perforated near the tip of the holder so that the stack could fit in there. The samples were secured in that position using a fine layer of kapton foil. The samples were tilted by 10$\degree$ with respect to the beam axis in order to reduce the interference with the measurements by the TOF detectors located in the forward direction at 0$^\circ$ in a farther position. Given the close distance between the neutron source and the samples, the difference in solid angle covered by the first and last samples in the stack was not negligible. This effect was also incorporated into the unfolding by correcting the activation depending on their distance to the source.

\begin{table}[h!]
\caption{List of samples and reactions used in the activation measurement setup. Cd-A and Cd-B stand for the cadmium thermal neutron covers. The ordering in the table represents the actual arrangement of the samples in upstream-to-downstream configuration. }
\label{tab:listsamples}       
\begin{tabular}{llll}
\hline\noalign{\smallskip}
 Sample & Mass (g) & Reactions & Half-life \\ 
 \noalign{\smallskip}\hline\noalign{\smallskip}

Au-a & 0.25874(2) &	$^{197}$Au(n,$\gamma$)$^{198}$Au & 2.6941 d\\
Cd-A & 1.13244(3) & -\\
Au-b & 0.26644(5) &	$^{197}$Au(n,$\gamma$)$^{198}$Au  & 2.6941 d \\&&$^{197}$Au(n,2n)$^{196}$Au & 6.1669 d \\&&$^{197}$Au(n,4n)$^{194}$Au  & 38.02 h \\
Al   & 0.06468(5) &	$^{27}$Al(n,$\alpha$)$^{24}$Na  & 14.997 h\\
In   & 0.26002(1) &	$^{113}$In(n,3n)$^{111}$In  & 2.8047 d\\&& $^{113}$In(n,4n)$^{110}$In  & 4.92 h\\ &&$^{115}$In(n,$\gamma$)$^{116m}$In  & 54.29 min\\ && $^{115}$In(n,n')$^{115m}$In  & 4.486 h\\
Ti  & 0.28573(3) &	$^{46}$Ti(n,p)$^{46}$Sc  & 83.79 d\\ && $^{48}$Ti(n,p)$^{48}$Sc  & 43.71 d\\

Co   & 0.31926(8) &	$^{59}$Co(n,2n)$^{58}$Co  & 70.86 d \\&& $^{59}$Co(n,4n)$^{56}$Co  & 77.236 d\\&&$^{59}$Co(n,p)$^{59}$Fe  & 44.49 d \\&&$^{59}$Co(n,$\alpha$)$^{56}$Mn  & 2.5789 h \\
Ni   & 0.57104(4) &	$^{58}$Ni(n,p)$^{58}$Co  & 70.86 d \\&&$^{58}$Ni(n,2n)$^{57}$Ni  & 35.6 h\\
Au-c & 0.24924(2) &	-  \\
Cd-B & 0.93009(2) & -\\

\noalign{\smallskip}\hline
\end{tabular}
\end{table}

\subsubsection{Activation measurements}

\begin{figure}
  \centering\includegraphics[width=0.45\textwidth, angle=-90]{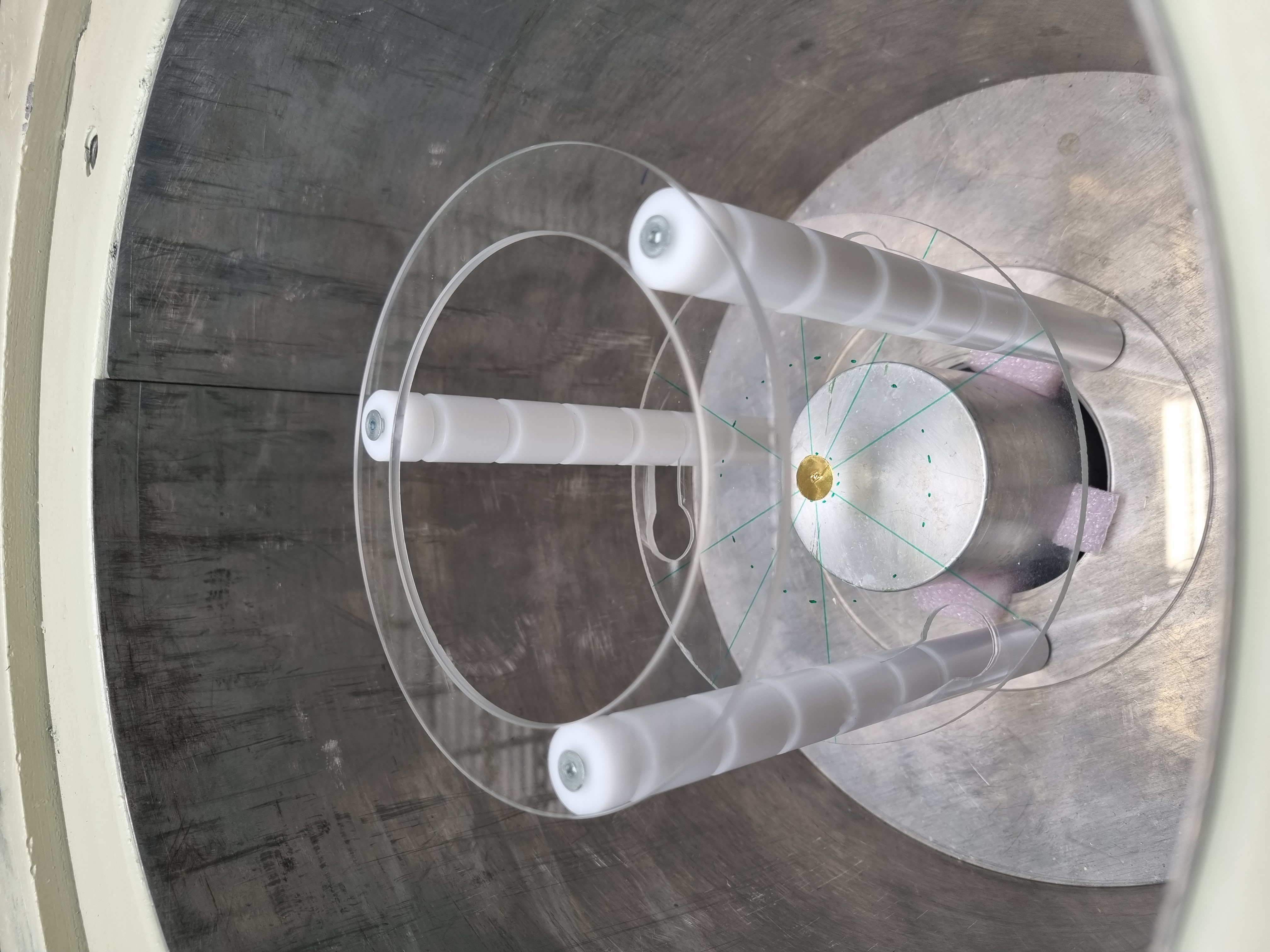}
\caption{Gamma spectroscopy station with the HPGe detector and the surrounding lead shielding. A thin methacrylate support was used to position the samples. A gold foil is shown in the sample position at 3 cm from the detector.}
\label{fig:HPGeAuSample}       
\end{figure}

The activity of each isotope produced by irradiation of the samples was measured using Gamma spectroscopy with an HPGe detector.
It was placed inside a lead well and cooled every 12 hours with liquid nitrogen. A shaping amplifier ORTEC 672 was used and pulses were digitized with an ORTEC 928 MCB and the MAESTRO 7.01 DAQ.
The efficiency $\varepsilon(E_\gamma)$ of the HPGe detector was calibrated with several certificated radioactive gamma sources (\tssc{22}Na, \tssc{57}Co, \tssc{60}Co, \tssc{88}Y, \tssc{133}Ba and \tssc{137}Cs) for that purpose. Several positions at different distances (3-15 cm) between the detector and samples were used to keep a balance between detection efficiency and dead-time losses. A dedicated low perturbating sample holder has been designed and 3D printed to guarantee the accuracy of the efficiency calibration and measurements. Simulations were carried out to account for the different sizes of the calibration radioactive sources and the measured irradiated samples, obtaining correction factors to the efficiency, $\kappa_\varepsilon$. The low activity of the irradiated samples impelled in most cases to close the distance down to 3 cm to increase the count rates. Even if generally small, dead time corrections were taken into account. The self-shielding correction factor, $\kappa_s$, was also computed in dedicated simulations for every gamma line of each sample. Figure \ref{fig:HPGeAuSample} shows the setup of the gamma spectroscopy station.

\begin{figure}
  \includegraphics[width=0.50\textwidth]{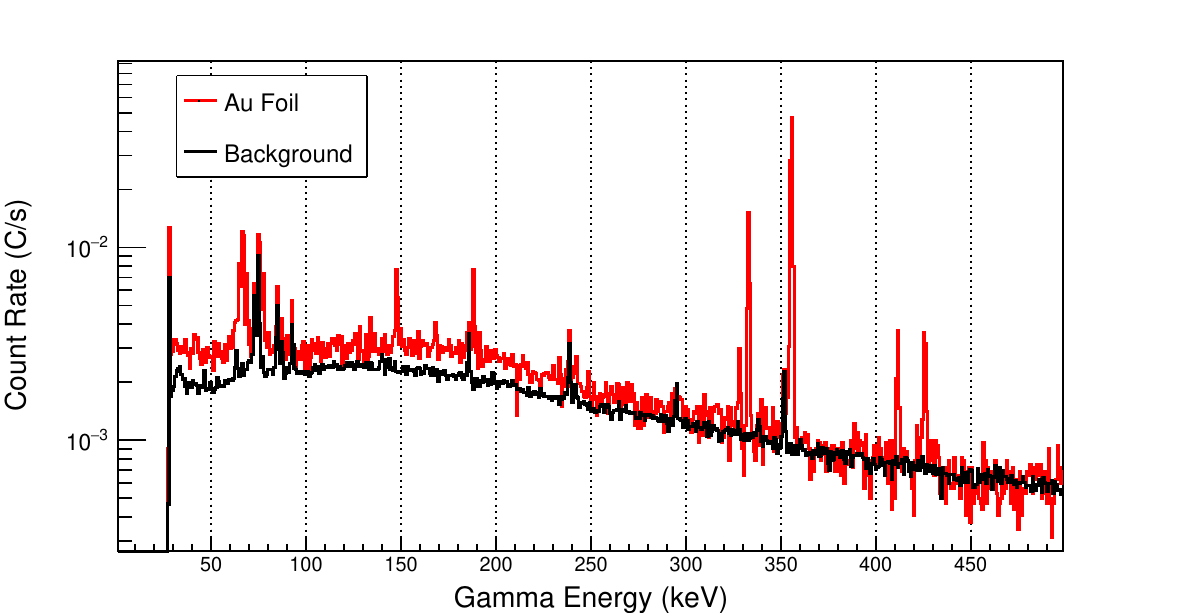}
  \includegraphics[width=0.50\textwidth]{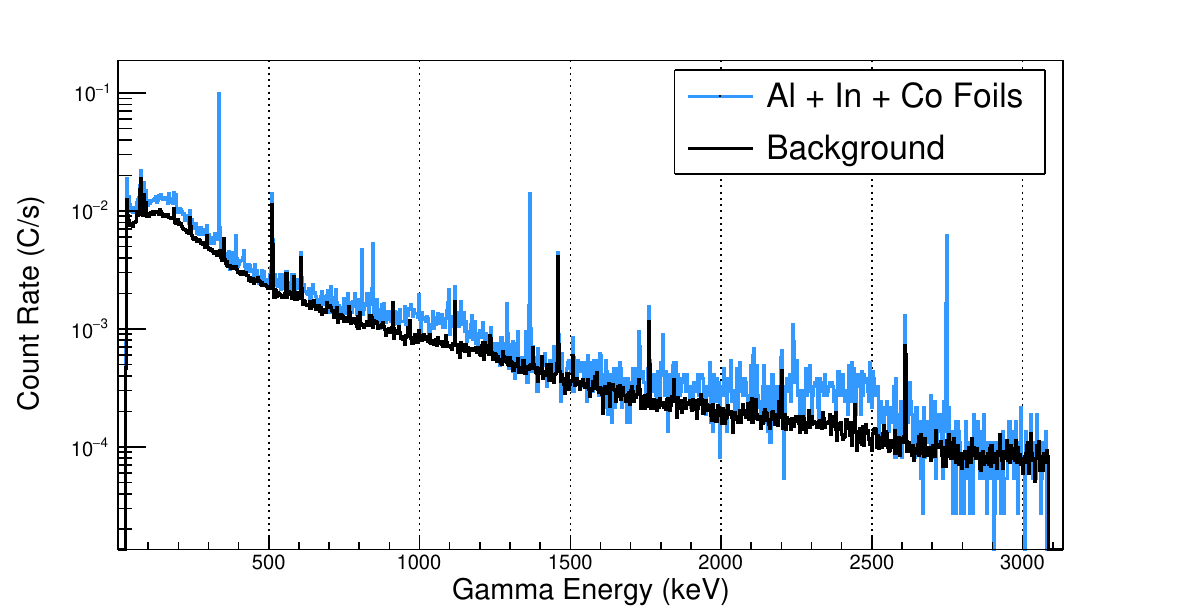}
\caption{\textit{(Top)}: Gamma spectrum of the Au sample compared to the overall background.\\
\textit{(Bottom)}: Gamma spectrum of the joint measurement of the Al, In, and Co samples compared to the overall background.}
\label{fig:GammaSpectra}       
\end{figure}

After the end of irradiation, a schedule of measurements was followed in order to obtain the activity data with the best precision, given the time constraints. Special regards had to be considered given the relatively short half-lives of some of the produced isotopes, which were assessed by measuring some of the samples jointly (e.g., Al (containing $^{24}$Na), In (containing $^{110}$In, $^{115m}$In and $^{116m}$In) and Co (containing $^{56}$Mn). Figure \ref{fig:GammaSpectra} shows some of the measured spectra for these samples.

The samples were measured after some cooling time, $t_c$, from the end of irradiation (EOI). The activity of isotope $k$ at the end of irradiation, $A_{k}^{EOI}$, was extrapolated from the measured counts in a certain gamma line, $C_\gamma$, during the measuring time, $t_m$ using the following expression:
\begin{equation}
    A_{k}^{EOI} = \frac{C_\gamma}{n_\gamma\varepsilon\cdot\kappa_\varepsilon \cdot \kappa_s}\cdot \frac{\lambda_k}{1-e^{-\lambda_k t_m}}e^{\lambda_k t_w} 
\end{equation}
where energy $E_\gamma$, relative intensities of the gamma rays $n_\gamma$, and the decay constants $\lambda_k$ of the produced isotopes were taken from ENSDF \cite{ENSDF}. At least one gamma line was used for each reaction. More than one was employed when available. An average value between the activity values from each gamma line was used in these cases.

\subsubsection{Unfolding routine}

In order to recover the spectral information from the activity of the samples, an unfolding procedure has to be performed. In the case of a steady-state neutron field, the activation rate equals the decay rate after several half-lives of the produced isotope. Therefore the saturated activity, $A_S$, can be used directly as input to the unfolding ($A_S = n\sigma\dot{\Phi}$). However, if the neutron flux is not stable in time (or not enough compared to the half-lives of the isotopes of interest), a correction factor has to be applied. In this sense, instead of keep using the saturated activity outside of its true definition, the total number of activations per atom in the samples, $R/n$, can be used. This in turn can be done by integrating in time (using the information from the irradiation history). This leads to an activations-to-fluence relation in order to perform the deconvolution:
\begin{equation}
    \frac{R}{n} = \sigma\cdot\tilde{\Phi}
\end{equation}

In a real experiment, the fluence measured by a foil, $\tilde{\Phi}$, is not exactly the true fluence in that position, $\Phi$, due to multiple scattering and self-shielding effects (and other effects from the surrounding support and other samples in the case of a stack), therefore, a correction factor $\kappa$ has to be introduced to the response function, which is dominated by the cross-section:
\begin{equation}
    \frac{R}{n} = \sigma\cdot\kappa\cdot\Phi
\end{equation}
The energy spectrum can be subsequently divided into several energy groups, $\Phi_j$, that provide information on the shape of the spectrum and can be chosen depending on the relative contribution to the activations on sample $i$, $\left(\frac{R}{n}\right)_i$, mediated by the response function, $M$:

\begin{equation}
    \left(\frac{R}{n}\right)_i =\sum_j \sigma_{ij}\kappa_{ij}\Phi_j =\sum_j M_{ij}\Phi_j
\end{equation}

Under this approach, the response matrix includes an effective cross-section within each energy group, $\sigma_{ij}$, that has to be calculated and included in the response function as:

\begin{equation}
    M_{ij} = \frac{\int^{E_{j+1}}_{E_{j}} \sigma_i(E)\kappa_i(E) \Phi(E) dE }{\int^{E_{j+1}}_{E_{j}} \Phi(E) dE  }
\end{equation}

For the computation of the response function, cross-section data spanning from the low energy range up to above tens of MeV is needed. For this reason, the TENDL-19 extended cross-section evaluation database was used (including up to 60 MeV or 200 MeV depending on the isotope and reaction) \cite{tendl19}. In the cases where the cross-section data was not available above 60 MeV, linear extrapolation was applied.
These matrix elements are thus dependent on the intra-group shape of the neutron spectrum. The selection of any spectrum for the calculation of this matrix impacts the final unfolding, which should be chosen carefully. This impact can be assessed by performing variations in the guess spectrum and including that in the unfolded flux uncertainty.

In order to perform the unfolding of the spectrum, several algorithms are available, including iterative methods such as the SAND-II code \cite{SANDii}. Lately, there have been efforts to use other methods to manage the uncertainties more comprehensively. One such strategy uses  a Bayesian unfolding routine coupled to Markov Chain Monte Carlo simulations (MCMC) using the code JAGS \cite{JAGS}. This methodology is based on the one presented by Chiesa \textit{et al.} \cite{ChiesaBatman2,ChiesaBaTMAN}. Different sources of uncertainty can be handled, including the uncertainty in the experimental data from the activation of the samples, the aforementioned uncertainty in the spectrum shape and the uncertainties in the cross-section data, which prove to be very relevant in the MeV energy range. This approach has recently been used to unfold the spectrum of the newly built NEAR station of n\_TOF at CERN~\cite{NEAR_ND2022}.

\subsection{The TOF technique}

To provide a complementary and separate measurement of the spectral shape with higher energy resolution a fast liquid scintillation detector was deployed in the neutron beam to allow TOF measurements, especially for the high energy part of the spectrum. In the following paragraph, both the detector used and the associated electronics are briefly introduced.

\subsubsection{Experimental setup}



\begin{figure}[hbt!]
  \includegraphics[width=0.48\textwidth]{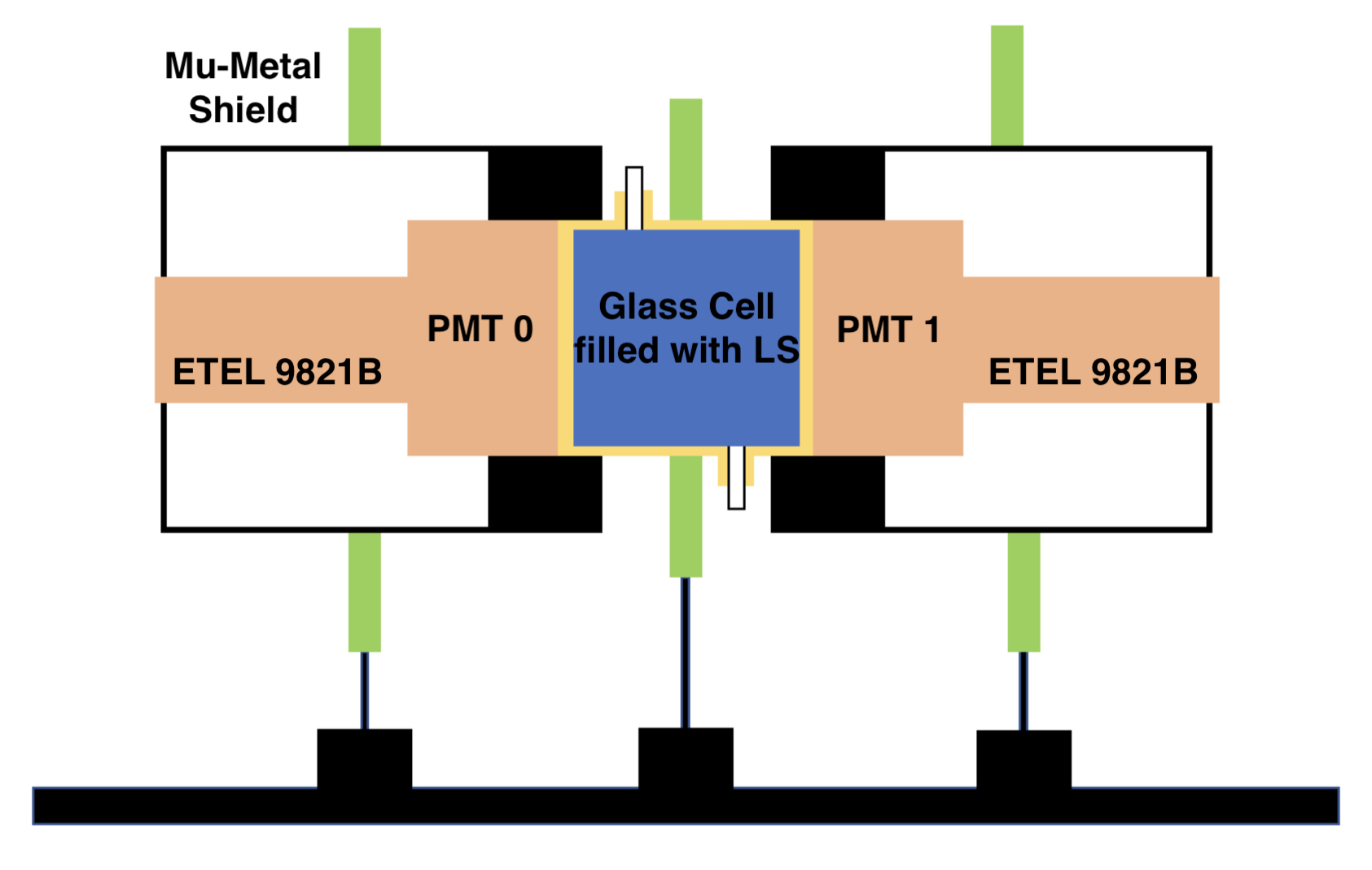}
\caption{Conceptual drawing of the experimental TOF setup consisting of: cylindrical glass cell filled with an LAB based liquid scintillator (LS), two  ETEL 9821B photomultiplier tubes (PMTs), the whole setup is surrounded by a mu-metal shield. 
}
\label{fig:TOF_Setup}       
\end{figure}

A custom-made liquid scintillator detector was used for the measurements presented here. Figure \ref{fig:TOF_Setup} shows a conceptual drawing of of this device. A 3x3 inch cylindrical cell of borosilicate glass with 1 mm thin optical windows was filled with a Linear alkylbenzene (LAB) based scintillation cocktail with admixtures of 2.5 g/l PPO (2,5-Diphenyloxazole) and 3mg/l BisMSB (1,4-bis-(o-methylstyryl)-benzene). LAB was chosen as the detector medium for safety reasons, due to its low toxicity and high flashpoint ($\sim$ 140 $^{\circ}$C). To protect the scintillator from oxygen, it was previously flushed extensively with high-purity nitrogen. After filling, the glass cell was sealed gas-tight under moderate overpressure (a few mbar) of nitrogen. To enhance the light collection efficiency the mantle surface of the cylindrical cell was previously covered by a directly evaporated aluminum mirror protected by a layer of plasma polymer. This vessel is coupled to two fast 3 inch 9821B photomultiplier tubes (PMTs) housed in mu-metal shields provided by ET Enterprises \cite{pmt}. The detector was housed in a thin-walled, fully enclosed aluminum dark box that serves as both a light shield and a Faraday cage against electromagnetic noise. This assembly was placed in the beam axis in a distance of 4.76 m behind the Li-target (distance with respect to liquid scintillators (LS) target cell center).

\begin{figure}[hbt!]
  \includegraphics[width=0.48\textwidth]{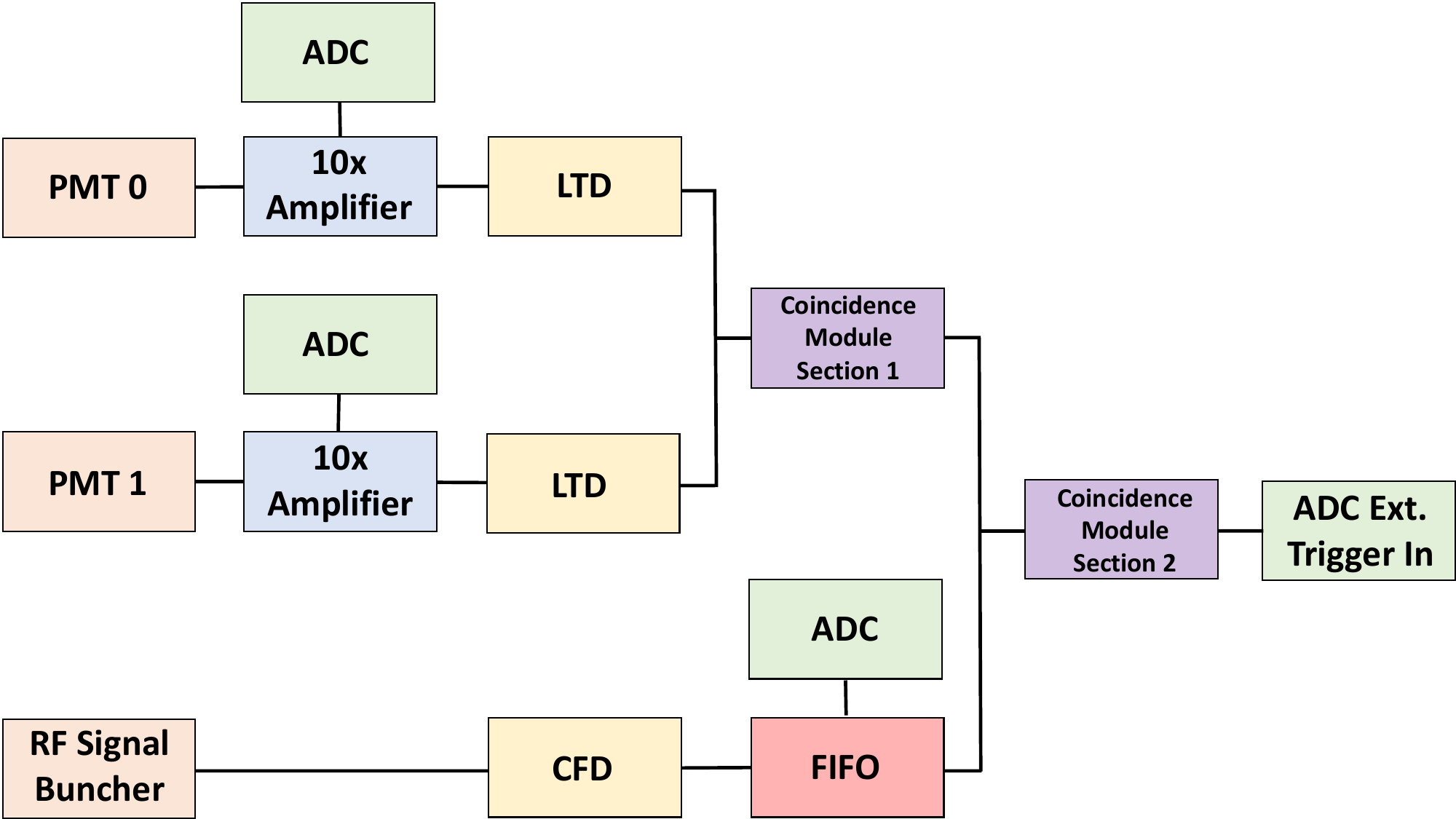}
\caption{Block diagram of the readout electronics: Photomultiplier Tube (PMT), RF Signal Buncher, Analog to Digital Converter (ADC), Fan in Fan out (FIFO), Low Threshold Discriminator (LTD), Constant Fraction Discriminator (CFD), 2 Coincidence Modules. 
}
\label{fig:TOF_readout_electronics}       
\end{figure}

The readout electronics of the TOF setup is based on a staged coincidence circuit. The first coincidence stage is realized between the two PMT branches and the second between the output of the first stage and the discriminated RF signal of the accelerator's beam buncher. The block diagram of the used electronics can be found in Figure \ref{fig:TOF_readout_electronics}. For TOF measurements described here, the PMT signals are fed to a CAEN N978 10x Fast Amplifier \cite{fastamp}. While one of the two outputs of each stage is directly connected to the waveform digitizer Agilent U1065A Acqiris DC282 (ADC), the other output is connected to a CAEN N844 LTD (Low Threshold Discriminator) \cite{ltd}.

The sinusoidal RF Signal used for the XTU Tandem's Beam Buncher is discriminated on the falling edge of the negative half-wave by a CAEN N842 CFD (Constant Fraction Discriminator) \cite{cfd}. The resulting logic signal is multiplied by a CAEN N625 Fan-In Fan-Out \cite{n625fifo} and sent to the ADC on the one hand and to the second stage of the double coincidence module on the other hand. In case of a coincident event in both PMTs in coincidence with the buncher signal, the waveform digitizer is triggered. While the LS detector reaches an internal time resolution of $\sigma_{LS}$=(769$\pm$6)~ps for energy depositions of 1~MeV$_{ee}$, the buncher of the XTU tandem was capable of producing beam bunches with a nearly gaussian shape and a width of $\sigma_{Bunch}$ =(1.08$\pm$0.02)~ns. Therefore, the beam properties are fully dominating the time and with that the energy resolution of the TOF spectrum.


\subsubsection{Unfolding routine}

\begin{figure}[hbt!]
  \includegraphics[width=0.48\textwidth]{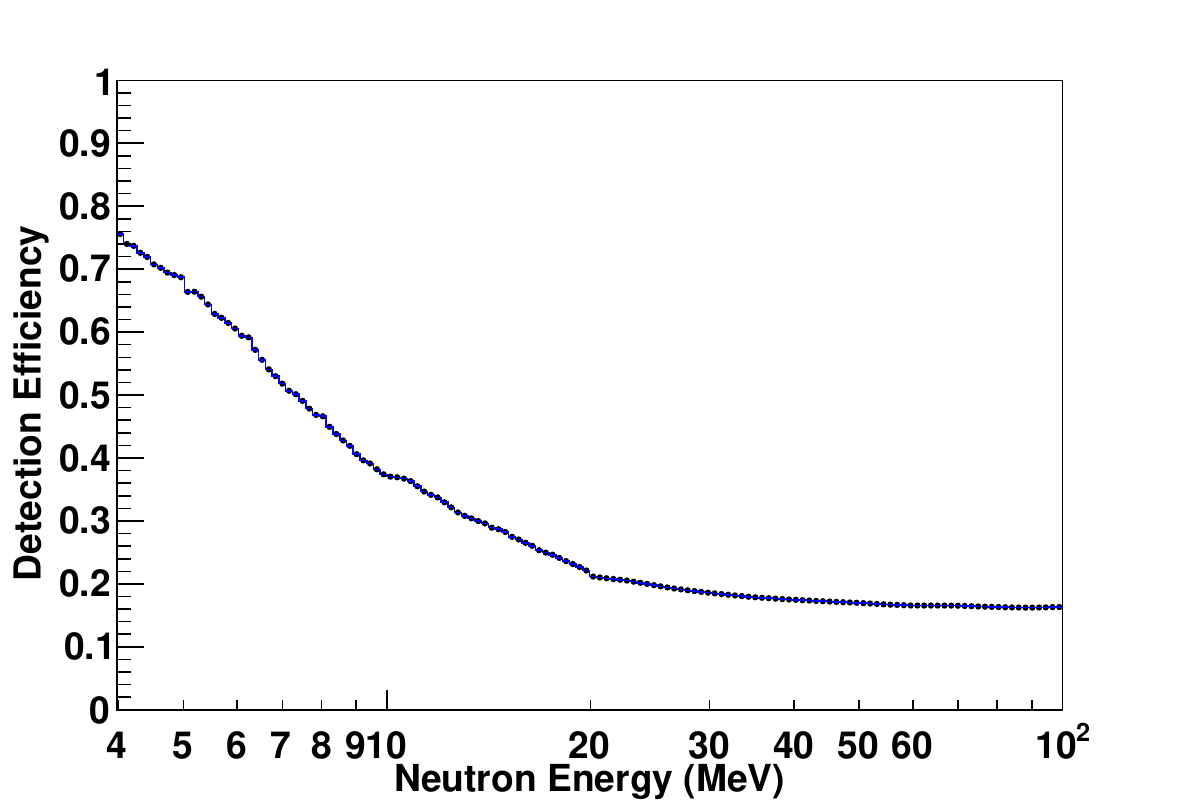}
\caption{Detection efficiency of the liquid scintillator located at 4.76 m from the neutron production target.
}
\label{fig:TOF_DetEff}       
\end{figure}

\begin{figure}[hbt!]
  \includegraphics[width=0.48\textwidth]{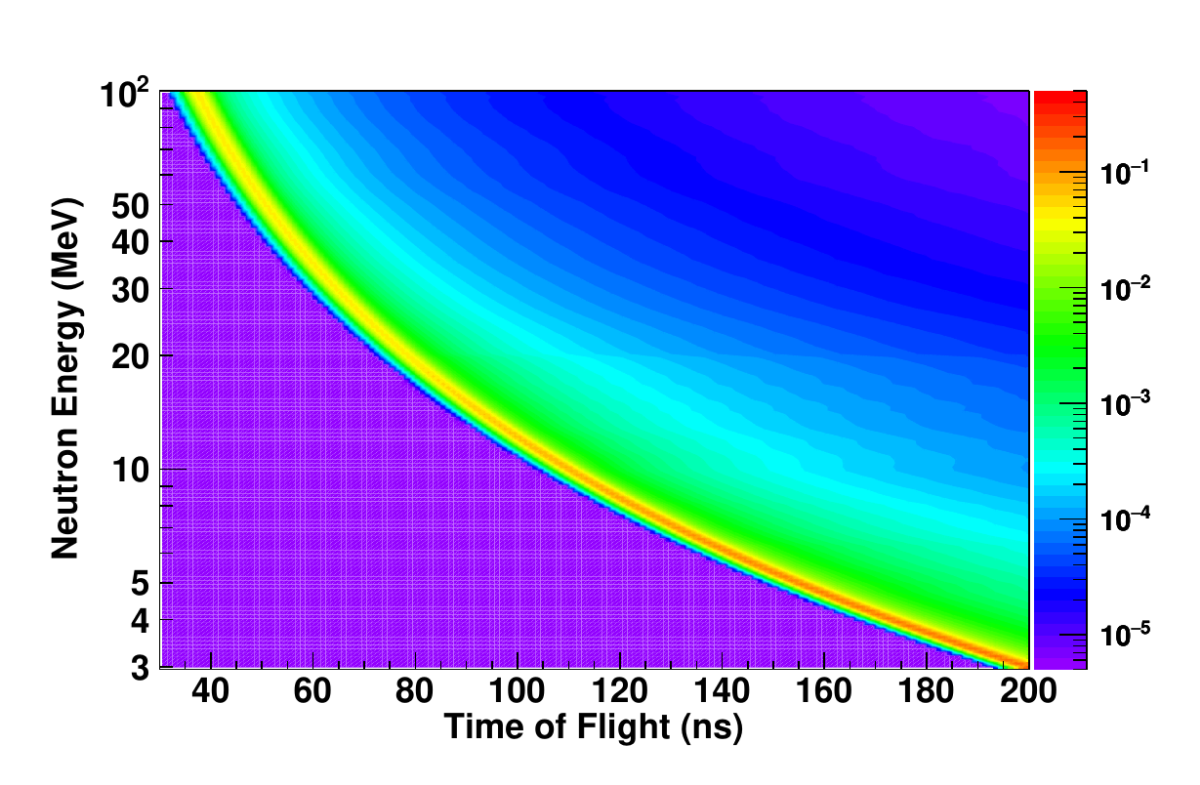}
\caption{Resolution function matrix used to convert TOF to neutron energy. The matrix includes the detection efficiency.}
\label{fig:RF}       
\end{figure}

The efficiency and the TOF-to-energy resolution function of the TOF detection setup were estimated from simulations with the MCNP6.2 code \cite{MCNP62}. Several simulations with monoenergetic neutron beams were carried out to compute the total detection efficiency, shown in Figure \ref{fig:TOF_DetEff} for the energy range covered in the experiment. Also time discretization in the experimental time window of 200 ns was included in the simulations in order to fill the resolution function matrix, as shown in Figure \ref{fig:RF}. The resolution function matrix provides a means to convert TOF distributions to neutron energy spectra, including the total flight path from the neutron source to the actual detection inside the liquid scintillator. Also the time width of the pulse was reproduced in the simulations by using the distribution observed in the detection of the $\gamma$-flash. For the calculations of the efficiency and resolution function, elastic scattering of neutrons off hydrogen and charged particle reactions from both hydrogen and carbon were included. Elastic scattering off carbon was not included due to the heavy quenching that prevented the detection of the carbon recoils.

The unfolding routine to convert TOF to energy spectra was a bayesian-based iterative algorithm using the maximum likelihood expectation-maximization (MLEM) method. This method has been previously used for comparable applications, for instance the unfolding of $\beta$-decay total absorption spectra \cite{Tain2007} and for $\gamma$-ray spectroscopy decomposition\cite{Matta2022}.


\section{Results}
\label{sec:resultsanddiscussion}

\subsection{Results from Activation and TOF measurements}

The activation produced during the irradiation was measured via Gamma spectroscopy, from which the activity at the end of irradiation, $A^{EOI}$, was inferred. Table \ref{tab:listActEOI} lists the  $A^{EOI}$ for each sample and reaction. Subsequent spectral results from the analysis using the activation unfolding method and the total neutron yield are shown in the following section, compared to the Fluka simulations.

%

%

\begin{table}[h!]
\caption{List of measured activity at end of irradiation, $A^{EOI}$, for each sample and reaction.}
\label{tab:listActEOI}       
\begin{tabular}{lll}
\hline\noalign{\smallskip}
 Sample & Reaction & $A^{EOI}$ (Bq) \\ 
 \noalign{\smallskip}\hline\noalign{\smallskip}

Au-a &	$^{197}$Au(n,$\gamma$)$^{198}$Au & $0.59\pm0.05$\\
Au-b &	$^{197}$Au(n,$\gamma$)$^{198}$Au  & $0.36\pm0.03$ \\&$^{197}$Au(n,2n)$^{196}$Au & $5.39\pm0.22$ \\&$^{197}$Au(n,4n)$^{194}$Au  & $0.57\pm0.05$ \\
Al    &	$^{27}$Al(n,$\alpha$)$^{24}$Na  & $4.0\pm0.4$\\
In   &	$^{113}$In(n,3n)$^{111}$In  & $0.051\pm0.004$\\& $^{113}$In(n,4n)$^{110}$In  & $0.056\pm0.020$\\ &$^{115}$In(n,$\gamma$)$^{116m}$In  & $3.8\pm0.6$\\ & $^{115}$In(n,n')$^{115m}$In  & $22.0\pm0.4$\\
Ti   &	$^{46}$Ti(n,p)$^{46}$Sc  & $0.117\pm0.019$\\ & $^{48}$Ti(n,p)$^{48}$Sc  & $1.72\pm0.05$\\

Co    &	$^{59}$Co(n,2n)$^{58}$Co  & $0.534\pm0.013$ \\& $^{59}$Co(n,4n)$^{56}$Co  & $0.107\pm0.006$\\&$^{59}$Co(n,p)$^{59}$Fe  & $0.113\pm0.009$ \\&$^{59}$Co(n,$\alpha$)$^{56}$Mn  & $2.05\pm0.16$ \\
Ni   &	$^{58}$Ni(n,p)$^{58}$Co  & $1.73\pm0.04$\\&$^{58}$Ni(n,2n)$^{57}$Ni  & $1.76\pm0.11$\\

\noalign{\smallskip}\hline
\end{tabular}
\end{table}

\begin{figure}[hbt!]
  \includegraphics[width=0.48\textwidth]{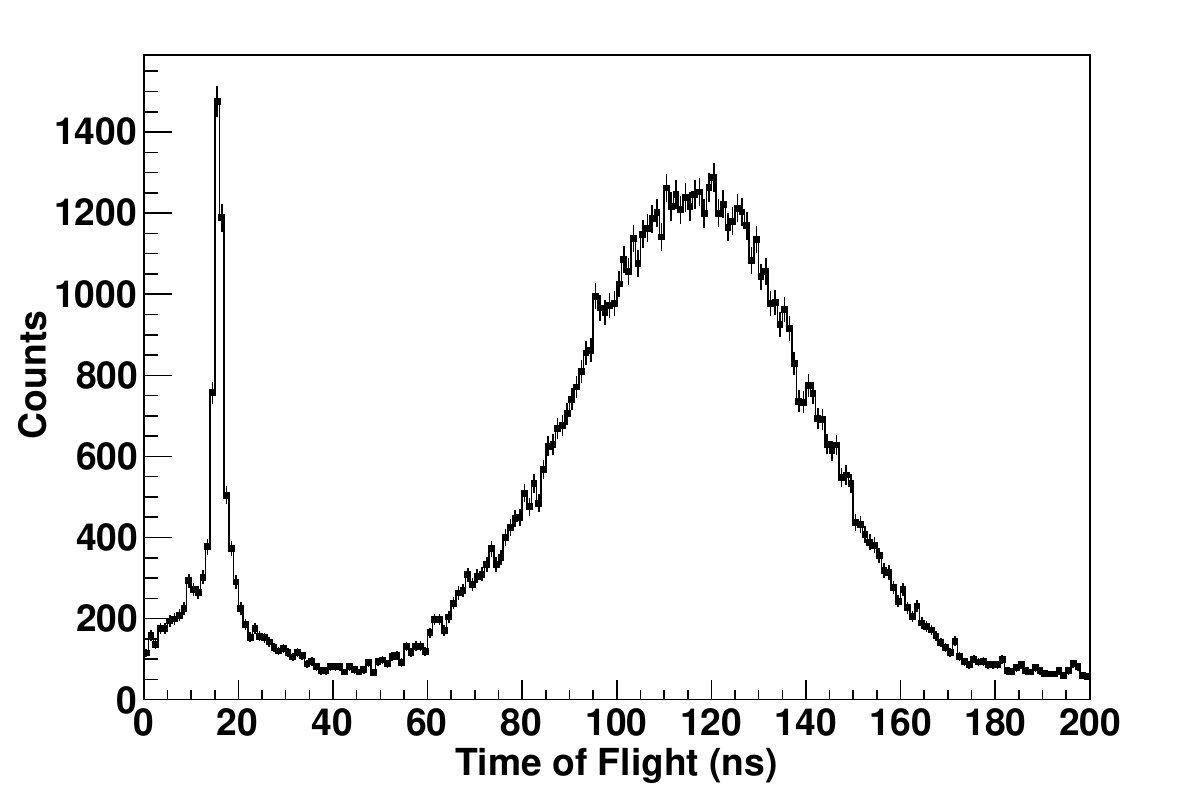}
\caption{Time of Flight data for events below 200 ns. The left peak shows the gamma flash events from the neutron production target, and the right peak shows the neutron distribution. From the gamma flash the time and width that energy resolution of the TOF spectrum was derived. The arbitrary offset for the TOF measurement is calibrated out by shifting the mean value of the gamma flash to the value derived from the division of the distance (between detector and Li target) by the speed of light.}
\label{fig:TOF_Data}       
\end{figure}

The experimental TOF spectrum measured with the LS detectors is shown in Figure \ref{fig:TOF_Data}. This TOF spectrum includes the $\gamma$-flash from the target and the neutron distribution. The long TOF tail is altered due to event rejection in triggering due to the fast bunch rate used in the experiment. This makes us reject the low energy tail in the reconstruction. Also this part of the experiment lacked a concise determination of the ion current used to produce the spectrum, and therefore only the shape of the spectrum is retained. The unfolded energy spectrum is shown in the next section.

\subsection{Comparison between FLUKA simulations and the experimental data}

\begin{figure}
  \includegraphics[width=0.48\textwidth]{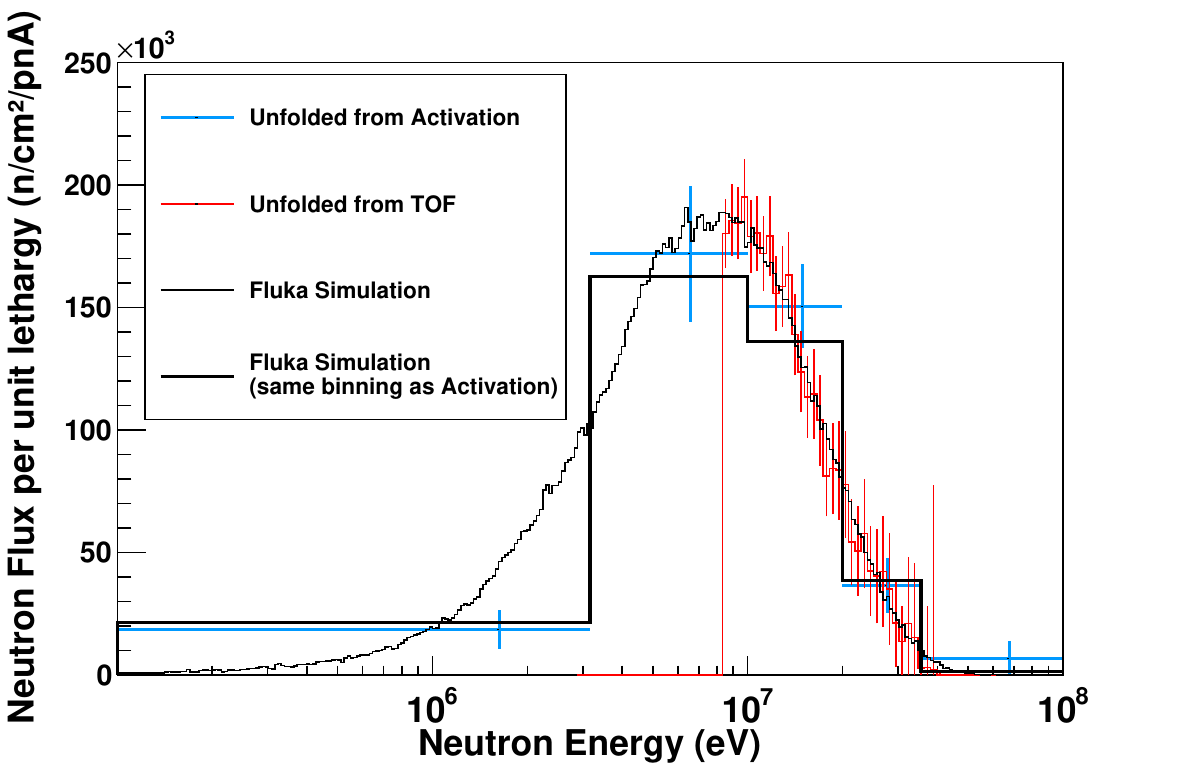}
\caption{Unfolded neutron spectrum obtained from activation and unfolded from TOF, compared with FLUKA simulations. FLUKA simulations are also shown with the same binning of the activation unfolding to facilitate the comparison.}
\label{fig:UnfoldingsvsFlukaSim}       
\end{figure}

Table \ref{tab:nflux} reports the measured neutron flux from unfolding at the activation foils position, compared with the FLUKA simulation; the unfolded neutron spectrum is shown in Figure \ref{fig:UnfoldingsvsFlukaSim}. The integral neutron flux of $(4.04\pm0.42) \times 10^{5}$ n/cm²/s/pnA at that position, confirms the estimates by FLUKA ( $(3.79\pm0.10) \times 10^{5}$ n/cm²/s/pnA), in good agreement within uncertainties.  The unfolded spectrum obtained from activation data is shown in blue in Figure \ref{fig:UnfoldingsvsFlukaSim}. Five energy groups above 0.1 MeV show good agreement both in shape and in the absolute value of the production yield.

\begin{table}[h!]
\caption{Measured neutron flux from unfolding at the activation position, compared with FLUKA simulation. The integral neutron flux is also given in the last row.}
\label{tab:nflux}       
\begin{tabular}{cccc}
\hline\noalign{\smallskip}
 \multicolumn{2}{c}{Energy Range (MeV)} & \multicolumn{2}{c}{$\Phi$ ($10^{3}$ n/cm²/pnA)} \\ 
 \noalign{\smallskip}\hline\noalign{\smallskip}
 Min & Max & Unfolded & FLUKA Sim \\
 \noalign{\smallskip}\hline\noalign{\smallskip}
$10^{-9}$ & $0.1$ & $11\pm35$ & $4.2\pm0.7$ \\ 
$0.1$ & $3.16$ & $64\pm27$ & $70.0\pm2.6$ \\
$3.16$ & $10.0$ & $198\pm32$ & $187.1\pm4.0$ \\
$10.0$ & $20.0$ & $104\pm12$ & $94.0\pm2.4$ \\
$20.0$ & $35.5$ & $21\pm6$ & $22.2\pm1.0$ \\
$35.5$ & $100$ & $7\pm7$ & $1.42\pm0.23$ \\
\noalign{\smallskip}\hline
$10^{-9}$ & $100$ & $404\pm42$ & $379\pm10$ \\
\noalign{\smallskip}\hline
\end{tabular}
\end{table}

Moreover, the unfolded data from the TOF measurement, covering the high energy part of the spectrum (above 8 MeV), shows a spectral shape with increased energy resolution and also in excellent agreement with the simulation. The results of the unfolding are shown also in Figure \ref{fig:UnfoldingsvsFlukaSim} in comparison with the activation and simulated data. The spectral data from TOF has been scaled to the activation data in that range as this part of the experiment did not include an independent normalization.


\section{Discussion and prospects}


A practical limitation of the $^7$Li($^{14}$N,xn)X
reaction at LNL is the intensity of the ion beam current, measured in particle nanoamperes compared to microamperes at the LNL proton machines. 
This limitation can be partially compensated for by placing test points just a few centimeters from the neutron production target. 

\begin{figure}[h]
  \includegraphics[width=0.48\textwidth]{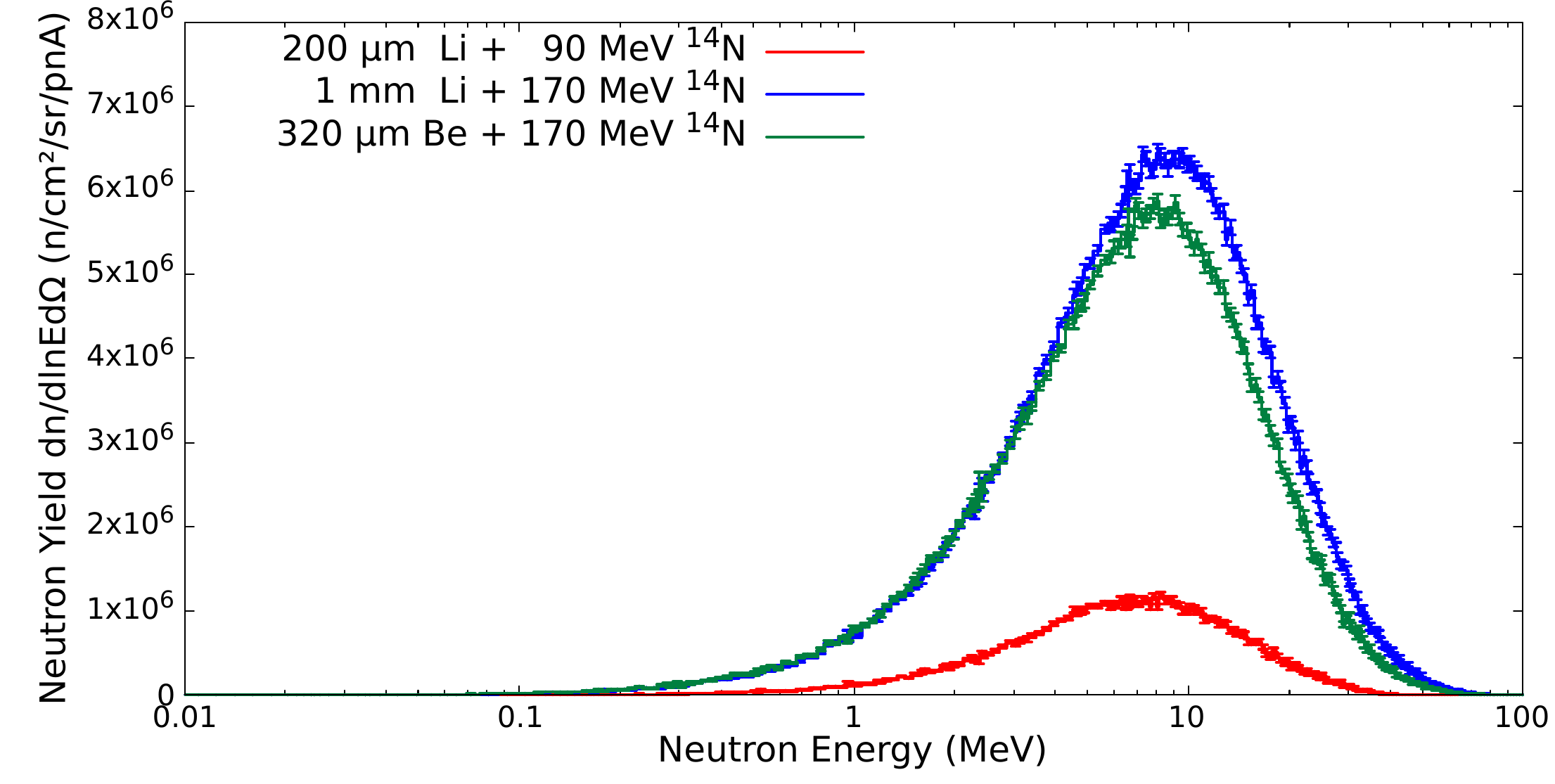}
\caption{FLUKA simulations comparing forward direction neutron spectra for 90 MeV energy using 200 $\mu$m with 170 MeV using 1 mm thick Lithium and 320 $\mu$m thick Be.}
\label{fig:ProspectsLiBe}       
\end{figure}

If only a Tandem is available,
the voltage and charge state, hence the beam energy, should be chosen to optimize the current and the energy dependent neutron yield. At LNL, the ALPI booster can be used to reach higher beam energies: 
according to FLUKA simulations, the energy spectrum can be stiffened and the neutron yield can be increased by a factor 5 by raising both the beam energy up to 170 MeV and the lithium thickness up to 1 mm (Figure \ref{fig:ProspectsLiBe}). 

In practice the transmission of the ALPI booster is only 8\%.
If the Tandem is used as an injector for ALPI, the gain in flux is more than canceled, but by using the more intense Electron Cyclotron Resonance (ECR) based PIAVE injection system for ALPI, the neutron flux can be significantly increased: the PIAVE system is capable of producing high positive charge state ions at the source, unlike the Tandem were an inefficient stripping foil must be used.
To verify the higher energy and thicker lithium simulations, we plan to perform the PIAVE-ALPI version of this experiment as soon as possible. 

Finally, we are investigating the possibility of using a thin beryllium easy to install self-standing foil target 
instead of the lithium targets fabricated in a glovebox (described in Section 2.1). Beryllium has
better physical properties (higher melting point, thermal conductivity, and lower linear thermal expansion coefficient) and is stable in air making it much more manageable compared to lithium. However the issue of radiation damage to the Be foil must be evaluated with great care: Be is toxic and the risk of polluting the beam line might be too great.


\section{Summary}
\label{sec:conclusions}
The $^7$Li($^{14}$N,xn)X reaction is cost effective and easy to implement at Tandem machines. LNL can use this reaction to immediately produce fast neutrons in an interesting energy range while waiting for the construction and commissioning of a dedicated neutron irradiation facility at the SPES proton cyclotron. The experimental neutron energy distribution measured is broad and peaked at 8 MeV, ranging from hundreds of keV up to tens of MeV. Activation foils  and TOF technique have been used to measure the produced energy spectra, showing a good agreement between them and the heavy ion fragmentation and coalescence models available in FLUKA. 

\begin{acknowledgements}
This work has been supported by the Cluster of Excellence PRISMA$^{+}$, the Cluster of Excellence ORIGINS as well as the Collaborative Research Center “Neutrinos and Dark Matter in Astro- and Particle Physics” (SFB 1258) and the University of Granada Chair "Neutrons for Medicine". Moreover, we are greatful for the support from the DFG Research Unit JUNO 2319 and 5519 (Project Number: 498394246) and the Spanish Junta de Andalucía (Projects FEDER no. A-FQM-371-UGR18 and B-FQM-156-UGR20). I.P. acknowledges finantial support from the Spanish Ministerio de Ciencia, Innovaci\'on y Universidades for his research stay at LNL under contract PRX18/00551. 
Furthermore, we would like to thank Francisco Ogallar Ruiz for his helpful advice and suggestions on the FLUKA simulations.

\end{acknowledgements}

\bibliographystyle{spphys}       
\bibliography{bib}   


\end{document}